\documentclass[english,preprint]{ipsj}

\usepackage{graphicx}
\usepackage{amsmath}
\usepackage{url}
\usepackage{comment}

\def\Underline{\setbox0\hbox\bgroup\let\\\endUnderline}
\def\endUnderline{\vphantom{y}\egroup\smash{\underline{\box0}}\\}
\def\|{\verb|}

\received{1111}{11}{11}
\accepted{1111}{11}{11}

\usepackage[varg]{txfonts}
\makeatletter%
\input{ot1txtt.fd}
\makeatother%

\definecolor{mygray}{gray}{0.6}
\definecolor{lg}{gray}{0.88}
\definecolor{pakistan}{rgb}{0.0, 0.5, 0.0}
\definecolor{kimidori}{rgb}{0.85,0.93,0.3}
\definecolor{mypink}{rgb}{0.9, 0.0, 0.4}
\definecolor{yamabuki}{rgb}{1.0, 0.86, 0.0}
\definecolor{navy}{rgb}{0.0,0.0,0.7}
\definecolor{darkred}{rgb}{0.7,0.0,0.0}

\newcommand{\temp}[1]{\colorbox{yamabuki}{#1}}

\begin{document}


\title{Machine learning approaches to explore important features behind bird flight modes}


\affiliate{WU}{Department of Pure and Applied Mathematics, Graduate School of Fundamental Science and Engineering, Waseda
University, Shinjuku, Tokyo, 169-8555 Japan}
\affiliate{TU}{Frontier Research Institute for Interdisciplinary Sciences, Tohoku University, Sendai, Miyagi 980-8578, Japan}
\affiliate{WUT}{Department of Applied Mathematics, Faculty of Fundamental Science and Engineering, Waseda University,
Shinjuku, Tokyo, 169-8555 Japan}

\author{Yukino Kawai}{WU}[yukino\_k@akane.waseda.jp]
\author{Tatsuya Hisada}{WU}[tatsuya34souhon@akane.waseda.jp]
\author{Kozue Shiomi}{TU}[shiomikozue@gmail.com]
\author{Momoko Hayamizu$^*$}{WUT}[hayamizu@waseda.jp \ ($^*$ \ Corresponding \ author)]

\begin{abstract}
Birds exhibit a variety of flight styles, primarily classified as flapping, which is characterized by rapid up-and-down wing movements, and soaring, which involves gliding with wings outstretched. Each species usually performs specific flight styles, and this has been argued in terms of morphological and physiological adaptation. However, it remains a challenge to evaluate the contribution of each factor to the difference in flight styles.  In this study, using phenotypic data from 635 migratory bird species, such as body mass, wing length, and breeding periods, we quantified the relative importance of each feature using Feature Importance and SHAP values, and used them to construct weighted L1 distance matrices and construct NJ trees. Comparison with traditional phylogenetic logistic regression revealed similarity in top-ranked features, but also differences in overall weight distributions and clustering patterns in NJ trees. Our results highlight the complexity of constructing a biologically useful distance matrix from correlated phenotypic features, while the complementary nature of these weighting methods suggests the potential utility of multi-faceted approaches to assessing feature contributions.

\end{abstract}


\maketitle

\section{Introduction}
\label{sec:introduction}
Birds have a variety of flight modes, such as flapping their wings rapidly up and down or soaring, i.e., gliding through the air with wings outspread \cite{hedenstrom1993migration}. The type of flight a bird uses is usually determined at species level, rather than by individual choice. For example, while larger species tend to be soaring types, not all species with the similar body mass engage in soaring flight. These species-specific flight modes are likely to be influenced by morphological, physiological, and ecological features. 

\begin{figure}[htb]
	\centering
	\includegraphics[width=8cm]{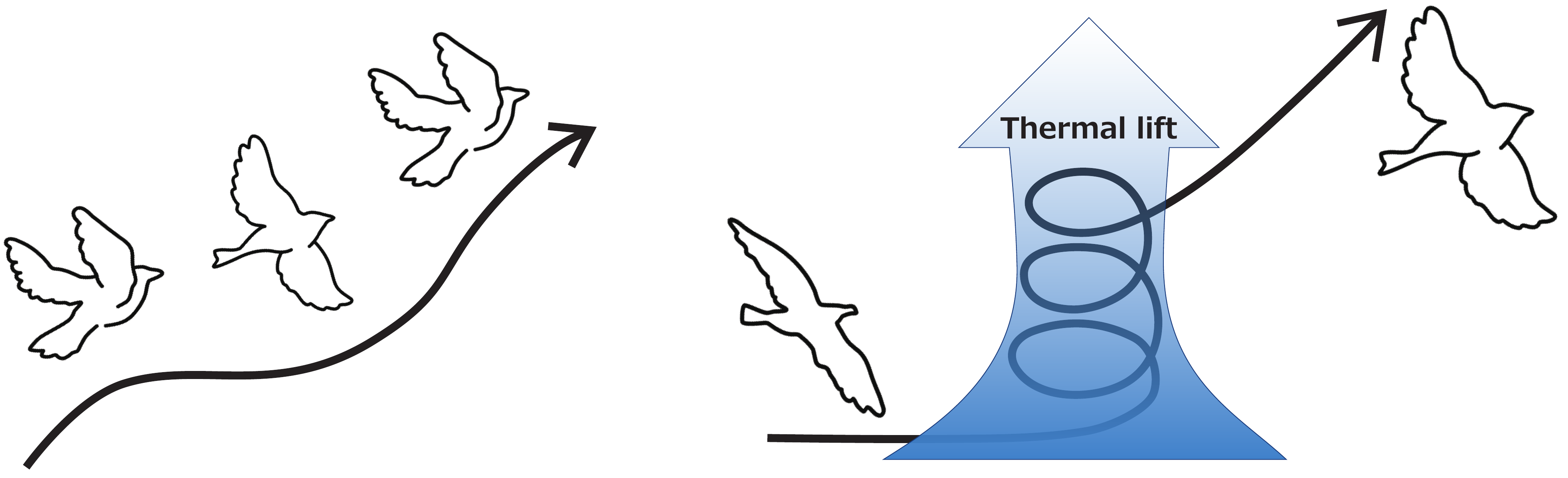}
	\caption{flapping type (left) and soaring type (right)}
	\label{figure:flyingtype}
\end{figure}

Several comparative studies have been conducted on birds regarding the morphological and ecological characteristics that are associated with the birds' flight mode. As the most important organ for flight, avian wings and wing-associated structures such as wing morphology, feather morphology, and bone strength, have been shown to correlate with flight modes \cite{akeda2023coracoid, kotnour2022relative, pennycuick2008modelling, tatani2024makes, taylor2014evolutionary}. In addition, it was found that soaring birds tend to have relatively larger brains \cite{published_papers/36140260}. This was explained by the energetic tradeoff hypothesis that assumes an energetic allocation to the brain and other organs \cite{aiello1995expensive, isler2006costs, isler2009expensive}. Less energetic cost in soaring flight could ease the energenetic limitation of the brain size.  Soaring species are also been known to spend longer time raising their young than flapping birds, which have been explained by their longer wings and larger brains requiring longer growth periods \cite{published_papers/36140260}.
These studies suggest that avian flight modes are not determined by a single feature, but by a complex interplay of multiple morphological, physiological, and ecological factors. However, the relative importance of each factor remains to be fully understood. While some studies examined the relative contribution of wing morphology or wing elements to discriminate flight modes \cite{shatkovska2017relationship, wang2011avian}, no integrative analysis including factors other than wing parameters has been performed.

For the purpose of estimating the contribution of factors involved in classification tasks, machine learning techniques such as FI (Feature Importance) and SHAP (SHapley Additive exPlanations) values are widely used.   These evaluation methods have been successfully applied in various fields including medicine and chemistry.
To date, however, no study has employed such machine learning approaches to quantify the contribution of various factors to the differences in bird flight modes. 

This paper provides a case study of application of machine learning to classification of bird species with different flight modes. We analyze publicly available phenotypic data of 635 migratory bird species and estimate the contribution of eight representative features such as body mass, wing length, and incubation period using two different approaches, FI and SHAP values. By employing these two quantification methods, we aim to identify the factors most closely associated with flight modes and to demonstrate the similarities and differences between the methods in the context of this analysis. To further clarify the differences, we not only obtained the FI and SHAP values, but also constructed distance matrices weighted by these values and compared the resulting Neighbor-Joining (NJ) trees.

This is the first study to apply machine learning techniques to quantify the relative importance of multiple features in determining avian flight modes. This study not only contributes to our understanding of avian flight mode evolution but also indicates the potential of machine learning approaches in quantifying the contribution of each phenotype to avian flight modes.

\section{Machine Leaning: FI and SHAP}

FI and SHAP values have been used in various areas of research. For example, in the prediction of polymer properties \cite{ehiro2023feature}, FI has been used to interpret the influence of different molecular features. In the classification of breast cancer MRI images \cite{vamvakas2022breast}, SHAP values were used to identify important image features.

While both FI and SHAP aim to quantify feature contributions, they differ significantly in their approach and interpretation. 
FI provides a global view of feature importance across the entire dataset, calculating importance based on how often each feature is used in decision trees and how much it improves classification performance. This method tends to emphasize features with overall impact, but may oversimplify complex interactions. 

In contrast, SHAP offers a more nuanced approach, with both global and local interpretations. It calculates the impact of each feature value on individual predictions relative to a base value, capturing complex feature interactions and nonlinear effects.

\section{Datasets}
\label{sec:datasets}
We utilized public phenotypic data on 635 migratory bird species (94 soaring and 541 flapping species) to investigate the relationship between morphological and ecological features  and flight mode. 
In this study, the flight mode of each species (whether soarer or flapper) was defined at the family level. More precisely, referring to previous studies  \cite{hedenstrom1998fast, simons2011cross, watanabe2016flight}, all species of the families Accipitridae, Anhingidae, Cathartidae, Ciconiidae, Diomedeidae, Falconidae, Fregatidae, Pelecanidae, and Procellariidae were labeld as soarers (dynamic soarers and thermal soarers), whereas all others were labeled as flappers.

We selected the following phenotypic features for the reasons we now explain. 

\begin{figure}[htb]
\centering
\includegraphics[width=3cm]{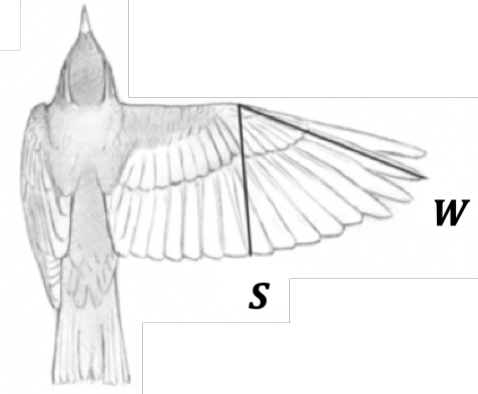}
\caption{Illustration of wing length $W$ and the distance $S$ from the carpal joint to the tip of the second primary feather  used to calculate HWI \cite{sheard2020ecological}.}
\label{fig:HWI}
\end{figure}

\begin{enumerate}
\item Body mass [g]
\item Wing length [mm]
\item Hand-Wing Index (HWI)
\item Beak length [mm]
\item Brain mass [g]
\item Tarsus length [mm]
\item Incubation period [$\log_{10}$ day]
\item Fledgling period [$\log_{10}$ day]
\end{enumerate}
Body mass was included as its relationship with soaring flight has been noted in earlier studies on bird flight modes.
Wing length and HWI are related to flight capabilities, as wings are the primary organs used for flight. HWI is a measure of wing slenderness, which is defined to be the ratio  
$100(W-S)/W$, 
where $W$ is the wing length and $S$ is the distance from the carpal joint to the tip of the second primary feather (Fig.~\ref{fig:HWI}). Brain mass, Incubation period, and Fledgling period were included based on known relationships between these features and flight modes \cite{published_papers/36140260} as described in Section \ref{sec:introduction}.

Tarsus length was included as a comparative feature, representing the hindlimb dimensions. A developmental trade-off between forelimb and hindlimb was demonstrated for birds \cite{heers2015wings}, and  it was shown that relative investment to forelimb and hindlimb was associated with flight modes \cite{kotnour2022relative}. By contrast, Beak length was included as a morphological feature that is not directly associated with flight modes.

The data for this study were compiled from several sources. Morphological measurements, including Wing length, HWI, Beak length, and Tarsus length, were obtained from AVONET \cite{tobias2022avonet}. Information on reproductive parameters, specifically Incubation period and  Fledging period, was gathered from studies by Cooney et al.~\cite{cooney2020ecology}, Minias et al.~\cite{minias2020avian}, and Jim{\'e}nez-Ortega et al.~\cite{jimenez2020long}. The remaining parameters were derived from the dataset provided by Shiomi \cite{published_papers/36140260}, where brain mass and body mass data were collected from various studies, including Sayol et al.~\cite{sayol2018predictable}, Ksepka et al.~\cite{ksepka2020tempo}, and Fristoe et al.~\cite{fristoe2017big}, and the flight mode classification was determined by consulting previous literature \cite{hedenstrom1998fast, simons2011cross, watanabe2016flight}.

\section{Experimental procedure}

\subsection{Calculation of FI and SHAP values}
We first standardized the data values for each feature listed in Section \ref{sec:datasets} for the 635 migratory bird species using a larger reference dataset of 2242 bird species (including both migratory and non-migratory birds) collected from multiple datasets mentioned in Section \ref{sec:datasets}. More precisely, for 2242 bird species, the values $x_i$ of each feature $i$ are standardized to $x_i^\prime:= (x-\overline{x_i})/\sigma_i^2$  so that the mean $\overline{x_i}$ and variance $\sigma_i^2$ of the $i$-th feature over 2242 bird species are $0$ and $1$, respectively. Using these standardized values, we represent each species using an $8$-dimensional feature vector. 
 
While decision tree-based methods such as LightGBM do not require data scaling, this standardization enhances the interpretability of data values by providing meaningful relative measurements within the context of global bird diversity. We note that applying standardization to training and test data collectively could cause data leakage and overestimate the model's generalization performance, but this potential data leakage during standardization is not a concern since our goal is not to build a model that generalizes well to unseen data.

To compute the FI and SHAP values of each feature, we used the LightGBM Python library \cite{ke2017lightgbm} to solve a binary classification problem for 635 migratory bird species, categorizing them into soaring and flapping types.  A stratified 10-fold cross-validation was used, with 90\%  of the data in each fold as training data and the remaining 10\% as test data. Stratified cross-validation ensures that each fold contains approximately the same proportion of each flight style as the original dataset.
The LightGBM model was trained with the following parameters: boosting\_type: \texttt{gbdt} (gradient boosting decision tree), objective: \texttt{binary} (binary classification of soaring and flapping species), and random\_state: \texttt{0} (reproducible output).

During training, we checked the performance on both the training and validation sets, and stopped early if the validation scores did not improve over 100 rounds. The accuracy of the LightGBM model in binary classification (i.e., the proportion of correctly classified test samples) reached a mean of 0.9717 (SD: 0.0219) across 10-fold cross-validation.

While LightGBM offers two types of FI calculation functions, \texttt{split importance} based on the number of splits used in the model, and \texttt{gain importance} based on the information gain from the splits in the model, we calculated the FI values using \texttt{gain importance}  in this study. This is because \texttt{gain importance} can provide a more detailed evaluation of feature importance than the evaluation by the number of splits \cite{strobl2007bias}. The SHAP values were computed using the Python library \texttt{shap} \cite{shap}. The calculation used the trained LightGBM model and the explanatory variables (the eight features in Section \ref{sec:datasets}) as inputs.

For the subsequent analysis in Section \ref{sec:dist}, we computed the L1-normalized average FI and SHAP values of each feature. Namely, based on the 10-fold cross-validation, we first computed two eight-dimensional vectors representing the mean FI values and SHAP values, respectively. Then, each vector was L1-normalized to ensure that the sum of the eight values of the explanatory variables equals 1, allowing for direct comparison between FI-based and SHAP-based weights.

\subsection{Phylogenetic logistic regression}
For the purpose of comparison, we applied  phylogenetic logistic regression, which  extends standard logistic regression by accounting for phylogenetic non-independence between species \cite{ives2010phylogenetic, tung2014linear}  and has been used in avian research (e.g. \cite{phyloLogistic2020}), to the same standardized dataset. Indeed, logistic regression analysis yields a coefficient value $\beta$ for each variable, which can be seen as the relevance of that feature to the binary classification task. However, unlike the LightGBM model, logistic regression is essentially a linear model, so it may not capture complex nonlinear relationships between flight modes and morphological features. Also, unlike FI and SHAP values, the coefficient value $\beta$ can be positive or negative, so the interpretation of each feature's contribution may not be straightforward, particularly when there are correlations between the variables. Therefore, it is meaningful to see the similarity and difference between machine learning-based  approach and conventional one.

Phylogenetic logistic regression was performed using R (R Core Team 2023) with the function \texttt{phyloglm} in the package \texttt{phylolm} \cite{tung2014linear}. The dependent variable was flight mode, and predictor variables were the eight features in Section \ref{sec:datasets}. The statistical significance was assessed using Wald-type tests, conditional on the phylogenetic correlation parameter ($\alpha = 0.01004094$) although we will not use this information in this study.

The phylogenetic tree was based on the Kuhl-backbone tree \cite{published_papers/36140260}, which was created by grafting sub-clades from a maximum clade credibility tree \cite{cooney2017mega} onto the phylogeny backbone of Kuhl et al.~\cite{kuhl2021unbiased}. The original tree data was obtained from BirdTree.org \cite{jetz2012global, jetz2014global} using the Hackett backbone option \cite{hackett2008phylogenomic}.

\subsection{Computation of Distance Matrices and NJ trees}\label{sec:dist}
Using each of FI and SHAP values as weight coefficients for each feature, we computed two distance matrices $D_{\rm FI}$ and $D_{\rm SHAP}$ using the weighted $L1$ norm of the 8-dimensional feature vector. For comparison, we also computed an unweighted (uniformly weighted)  $L1$ distance matrix $D_{\rm ave}$ and a distance matrix $D_{\rm PL}$ based on the absolute values $|\beta|$ of the coefficient values $\beta$ obtained from phylogenetic logistic regression. Note that the computation of $D_{\rm PL}$ uses the coefficients  for all the eight features regardless of their statistical significance.

From each distance matrix, we constructed an NJ tree using the SplitsTree 6 software package (\url{https://husonlab.github.io/splitstree6/}). The resulting NJ trees were compared in terms of how soaring and flapping species clustered.

\section{Results}

\subsection{FI and SHAP values and their robustness}

\begin{figure}[htb]
	\centering
	\includegraphics[width=.42\textwidth]{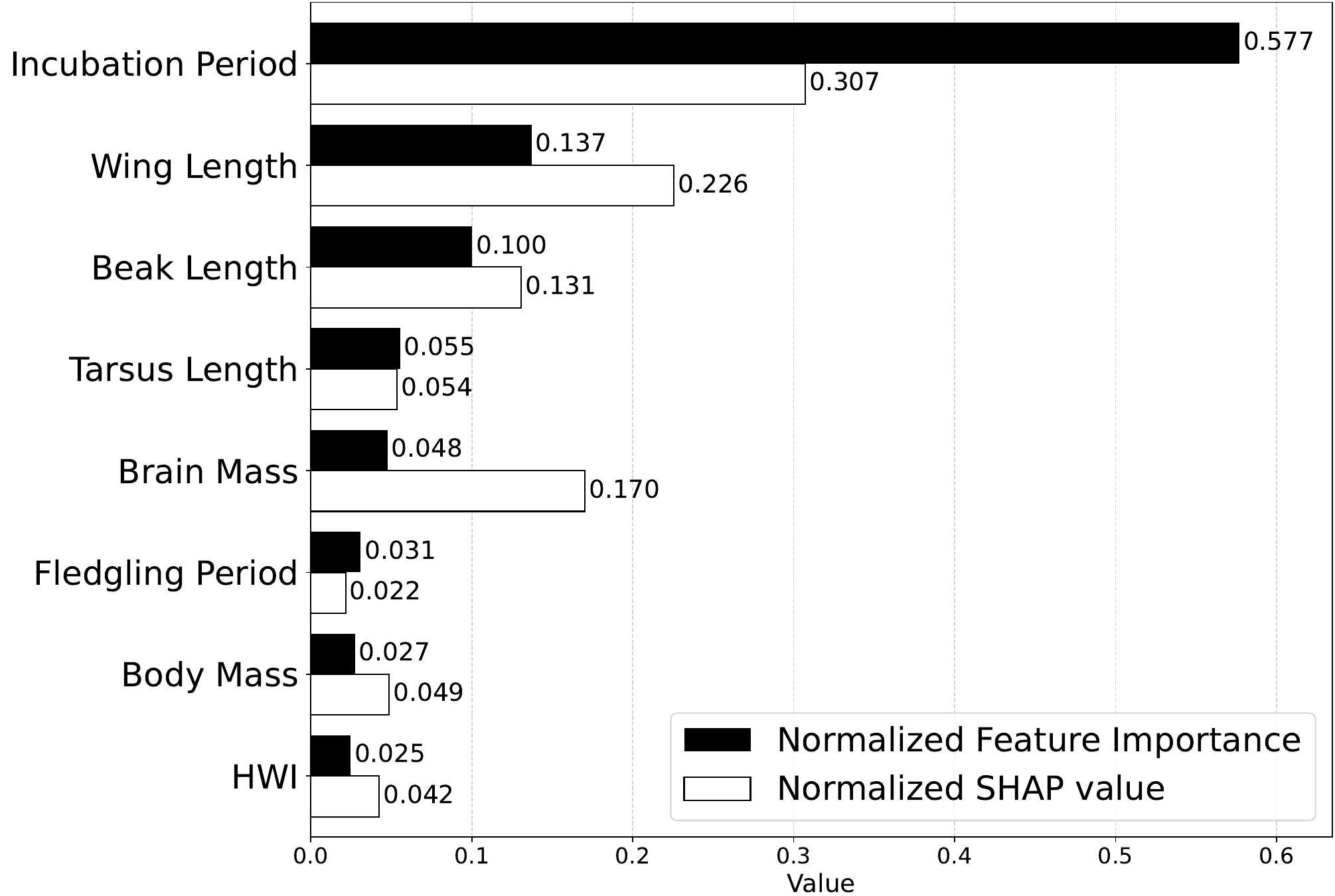}
	\caption{Normalized mean FI and SHAP values of the bird features}
	\label{fig:Normalized}
\end{figure}

We observed both similarity and difference between FI and SHAP results (Fig.~\ref{fig:Normalized}). The top two features were consistent: Incubation period was the greatest contributor, and Wing length was the second one. However, the weighting of the other features were different. For example, Beak length was ranked third by FI, but it was forth and Brain mass was third in the SHAP ranking. 

Another interesting difference found through cross-validation is the robustness of estimated values. As Table \ref{tab:CV} shows, FI had lower CV values than SHAP, meaning that estimated FI values are more stable than SHAP values. In fact, FI values of the top two features (Incubation period and Wing length) had CV below 0.1. In contrast, SHAP values  generally exhibited higher CV, with the CV value for Tarsus length reaching almost 0.96.

\begin{table}[htbp]
\centering
\caption{Unnormalized mean FI and SHAP values and their Coefficients of Variation (CV)}
\label{tab:CV}
\begin{tabular}{lrrrr}
\hline
Feature & FI Mean & FI CV & SHAP Mean & SHAP CV \\
\hline
Body mass & 66.924 & 0.286 & 0.357 & 0.760 \\
Wing length & 334.975 & 0.080 & 1.656 & 0.591 \\
HWI & 59.998 & 0.357 & 0.312 & 0.767 \\
Beak length & 244.763 & 0.091 & 0.959 & 0.451 \\
Brain mass & 116.181 & 0.224 & 1.250 & 0.574 \\
Tarsus length & 135.450 & 0.198 & 0.394 & 0.956 \\
Incubation period & 1409.940 & 0.026 & 2.256 & 0.394 \\
Fledgling period & 75.792 & 0.260 & 0.158 & 0.751 \\
\hline
\end{tabular}
\end{table}

\subsection{Results of phylogenetic logistic regression}
The phylogenetic logistic regression analysis also identified Wing length and Incubation period as the most relevant features. This is consistent with the FI and SHAP analyses, but their rank order was reversed: Wing length showed the strongest positive association ($\beta = 2.24, p < 0.001$), followed by Incubation period ($\beta = 1.77, p < 0.001$).
Coefficients for the other features differed substantially from machine learning approaches. Fledging period exhibited a significant negative association ($\beta = -1.11, p < 0.01$), which is unexpected given its positive correlation with Incubation period. The other features showed weaker associations: HWI ($\beta = -0.57, p = 0.14$), Body mass ($\beta = -0.42, p = 0.14$), Tarsus length ($\beta = -0.40, p = 0.17$), Brain mass ($\beta = -0.22, p = 0.59$), and Beak length ($\beta = 0.04, p = 0.86$). It is noteworthy that Beak length had the smallest absolute coefficient value because we expected Beak length to have no information to discriminate flight modes, as noted in Section \ref{sec:datasets}, but it did not rank lowest in the FI and SHAP analyses.

\subsection{Comparison of NJ trees}
\begin{figure*}[htb]
	\centering
	\includegraphics[width=\textwidth]{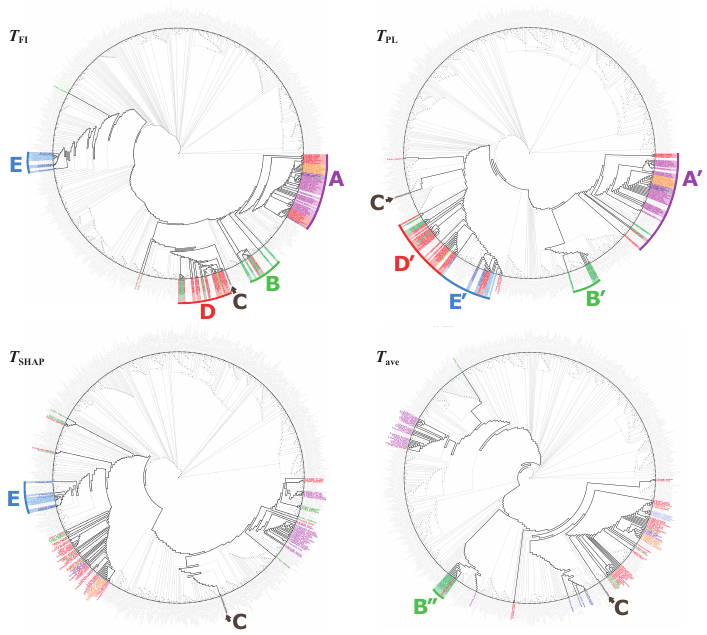}
	\caption{The NJ trees $T_{\rm ave}$, $T_{\rm FI}$, $D_{\rm SHAP}$, and $T_{\rm PL}$ constructed from the distance matrices $D_{\rm ave}$, $D_{\rm FI}$, $D_{\rm SHAP}$ and $D_{\rm PL}$. For ease of comparison, we have colored and annotated only the subtrees for the soaring species. A and A$^\prime$ are clusters mainly consisting of Accipitridae, Diomedeidae, Fregatidae, and Procellariidae families; B, B$^\prime$, and B$^{\prime\prime}$ are those mainly formed by Falconidae family; C is a singleton cluster of Anhingidae family; D and D$^\prime$ are clusters mainly consisting of Accipitridae, Cathartidae, and Falconidae families; E and E$^\prime$ are those containing all species of Ciconiidae and Pelecanidae families.}
\label{figure_4trees}
\end{figure*}

Fig.~\ref{figure_4trees} shows the NJ trees $T_{\rm ave}$, $T_{\rm FI}$, $T_{\rm SHAP}$, and $T_{\rm PL}$, where only the subtrees for soaring species are highlighted for visual clarity. These trees show substantial differences, demonstrating that the results of distance-based clustering are highly dependent on the weighting scheme of the phenotypic features.
Among these trees, only $T_{\rm FI}$ and $T_{\rm PL}$ show similarities in their structures and clustering patterns, although $T_{\rm FI}$ has been constructed without using any phylogenetic tree. Both trees have five clusters with comparable compositions of soaring species, such as clusters A and A$^\prime$ and D and D$^\prime$. Cluster E, containing all four species of Pelecanidae family (pale blue) and all five species of Ciconiidae family (blue), is present in both $T_{\rm FI}$ and $T_{\rm SHAP}$, while it appears as cluster E$^\prime$ with a modified composition in $T_{\rm PL}$.

\section{Discussion}\label{sec:discussion}
\begin{figure*}[htb]
	\centering
	\includegraphics[width=.95\textwidth]{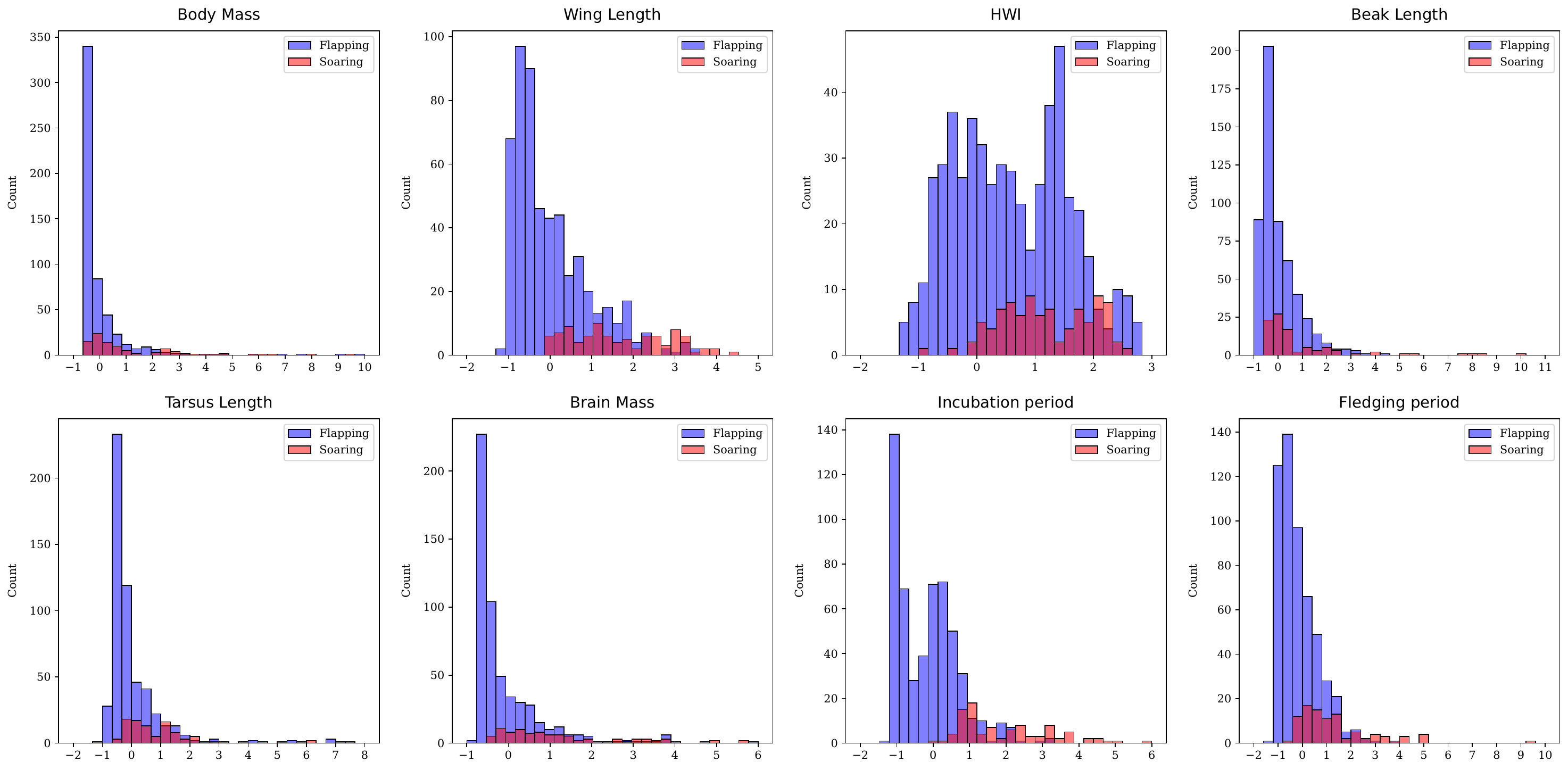}
	\caption{Histograms for the standardized values of each features in the flapping and soaring species.}
\label{figure_hist}
\end{figure*}

Our analysis revealed that all three methods (FI, SHAP, and phylogenetic logistic regression) identified Incubation period and Wing length as the two most important features for discriminating flight modes. The  importance of incubation period aligns with Shiomi  \cite{published_papers/36140260}. As Fig.~\ref{figure_hist} shows, the histograms for these features show less overlap between soarers and flappers, with soaring species generally showing higher values, supporting their high discriminative power. 

As we have seen in Fig.~\ref{fig:Normalized}, the FI and SHAP methods produced different distributions for the eight features. The distribution of FI values was skewed and the Incubation period received a remarkably high FI value. This can be attributed to the nature of FI, which calculates values based on the information contributing to the model splits. When multiple correlated features are present, which is the case in our data, FI tends to prioritize the feature with the strongest influence. By contrast, the SHAP results presented in Table \ref{figure_4trees} show a more balanced distribution of values compared to FI. SHAP evaluates the impact of the absence of certain features and, even in the presence of correlated features, can compensate for the absence of the strongest feature with other features, resulting in less biased values.

The histograms in Fig.~\ref{figure_hist} also help explain some unexpected weightings. For instance, Body mass and Brain mass, which have been associated with flight modes, did not emerge as relevant features in our analysis. This is not surprising because their histograms largely overlap between soarers and flappers in our dataset. Similarly, Beak length showed the weakest association in the phylogenetic logistic regression analysis while not in FI and SHAP analyses, but FI and SHAP may have detected that our dataset contains some soaring species with a longer beak.

We also note that none of the FI, SHAP, and the phylogenetic logistic regression analysis shows a causal relationship. For example, both the FI and SHAP methods agree on the importance of Incubation period, but this does not mean that the longer incubation period allowed some birds to evolve into soaring species.

\section{Conclusion}
In this paper, we applied machine learning techniques to quantify the contributions of avian phenotypic features to flight modes. 
By comparing FI and SHAP-based weighting approaches with traditional phylogenetic logistic regression, we found both consistencies and differences in their feature evaluations. 

All three methods identified the same top two contributing features, Incubation period and Wing length, which aligns with established biological knowledge. While FI assigned a significantly higher weight to Incubation period, SHAP provided a more balanced weight distribution across features. Such differences in the feature weights affected the resulting NJ trees. The robustness of FI values is also worth noting. The choice between these methods should be guided by specific research objectives and the nature of the data being analyzed.

Our approach demonstrates the potential of machine learning in estimating the contribution of phenotypic features and in constructing biologically meaningful distance matrices from phenotypic data. Future research could extend this methodology to a wider range of avian features and other biological classification problems, such as classification of sedentary or migratory bird species \cite{ives2010phylogenetic} and classification and feature selection using bird sound data \cite{LDAsound2019} or amphibian sound data \cite{LDA-tree-SVM2009}.

\begin{acknowledgment}
MH is supported by JST FOREST Program Grant Number JPMJFR2135. We also thank the anonymous reviewers for their careful reading and many useful comments.
\end{acknowledgment}

\bibliographystyle{ipsjsort-e}
\bibliography{kawai_2023_manuscript}

\end{document}